# Experimental Syntheses of Sodalite-like Clathrate $EuH_6$ and $EuH_9$ at Extreme Pressures

Liang Ma,[1,2] Guangtao Liu,[2] Yingying Wang,[2] Mi Zhou,[2] Hanyu Liu,[2] Feng Peng,[3] Hongbo Wang,[1,2*] and Yanming Ma[1,2,4†]

[1]State Key Laboratory of Superhard Materials, College of Physics, Jilin University, Changchun 130012, China

[2]International Center of Computational Method & Software, College of Physics, Jilin University, Changchun 130012, China

[3]College of Physics and Electronic Information, Luoyang Normal University, Luoyang 471022, China

[4]International Center of Future Science, Jilin University, Changchun 130012, China

**Corresponding authors:**
*whb2477@jlu.edu.cn, †mym@jlu.edu.cn

The recent discovery of a class of sodalite-like clathrate superhydrides (e.g., $YH_6$, $YH_9$, $ThH_9$, $ThH_{10}$, and $LaH_{10}$) at extreme pressures, which exhibit commonly a high-temperature superconductivity with the highest $T_c$ approaching 260 K for $LaH_{10}$, opened up a new era in search of high-temperature superconductors in metal superhydrides. There is a high interest towards the finding of alternative clathrate superhydrides that might witness the long-dreamed room-temperature superconductivity. Here, we target on the experimental synthesis of strongly-correlated europium (Eu) superhydrides where theory can fail for the prediction of superconductivity. We pressurized and laser-heated the mixture of metal Eu and ammonia borane ($NH_3BH_3$) in a diamond anvil cell and successfully synthesized the sodalite-like clathrate $EuH_6$ and $EuH_9$ at conditions of 152 GPa and 1,700 K, and 170 GPa and 2,800 K, respectively. Two non-clathrate structured phases of $EuH_5$ and $EuH_6$ were also synthesized that are not reported in lanthanide superhydrides. Calculated large H-derived electronic density of states at the Fermi level in clathrate $EuH_6$ implies the potential of high temperature superconductivity. Our work created a model superhydride platform for subsequent investigation on how strongly-correlated effect in electronic structure can affect the superconductivity of superhydrides, a phenomenon that is not known thus far.

## INTRODUCTION

In the year of 2012, there was a theoretical proposal on potential high $T_c$ superconductivity in superhydride $CaH_6$ (a theoretical $T_c$ = 235 K at 150 GPa) stabilized at high pressures [1]. The key to the unusually high $T_c$ superconductivity is the formation of a hydrogen (H) sodalite-like clathrate structure containing enclathrated Ca in a crystal lattice, giving rise to the large H-derived electronic density of states (DOS) at Fermi level and the extremely strong electron-phonon coupling related to H-H vibrations in the H cages [1]. The formation of a clathrate structure in group IV elements is common since the elements' four valence electrons are ready for acceptance of four covalent bonds to stabilize the clathrate cage [2-6]. However, the formation of such a clathrate strucgure for hydrogen that contains only one valence electron is quite unusual, and becomes possible only when hydrogen accepts sufficient extra electrons from Ca atoms under high pressure conditions [1].

Following the first prediction of $CaH_6$, the same sodalite-like clathrate structure was later proposed in $YH_6$ and $MgH_6$ with predicted $T_c$ = 264 K at 120 GPa [7] and $T_c$ = 260 K at 300 GPa [8], respectively. These two latter superhydrides together with $CaH_6$ share a common feature of high temperature superconductivity with theoretical $T_c$ values all exceeding 200 K. This appearance inevitably generated a great deal of attention towards finding of high $T_c$ superconductors in clathrate superhydrides.

In the year of 2017, along the line of finding clathrate structures in superhydrides, a major theoretical progress was achieved via a comprehensive crystal structure searching simulation on rare-earth (RE) superhydrides at high pressures [9]. The results unravel that pressure-induced formation of sodalite-like clathrate structures is a general behavior for all RE superhydrides. Besides $REH_6$ that shares the same clathrate structure with that of $CaH_6$, other two clathrate stoichometries of $REH_9$ and $REH_{10}$ having higher H content than that of $CaH_6$ were reported. Though all of them are clathrate structures, there is emergence of $H_{24}$, $H_{29}$, and $H_{32}$ cages for stoichiometries $REH_6$, $REH_9$ and $REH_{10}$, respectively. Notably, high temperature superconductivity appears frequently in these RE superhydrides with the predicted $T_c$ = 276 K at 150 GPa for $YH_9$, $T_c$ = 303 K at 400 GPa for $YH_{10}$, and $T_c$ = 288 K at 200 GPa for $LaH_{10}$ [9]. It should be particularly emphasized that the predicted clathrate structure and high temperature superconductivity of $LaH_{10}$ and $YH_{10}$ in Ref. [9] coincide exactly with the results from another independent theoretical work [10] appearing at nearly the same time.

Motivated by the theoretical prediction, experimental progresses on syntheses of these clathrate superhydrides are remarkable [11-19]. A major move was the observation of near-room temperature superconductivity in the clathrate LaH$_{10}$ with $T_c$ approaching 260 K [12,13]. Subsequent syntheses of clathrate superhydrides of YH$_6$ [15,16], YH$_9$ [15], ThH$_9$ and ThH$_{10}$ [14] with the observed high $T_c$ values of 227 K at 237 GPa, 243 K at 201 GPa, 146 K at 170 GPa and 159 K at 174 GPa, respectively, add more examples into the family of clathrate superhydrides, a new class of high temperature superconductors holding a record high $T_c$ value of 260 K among all known superconductors. The clathrate superhydrides of CeH$_9$, PrH$_9$, and NdH$_9$ as predicted in Ref. [9] were also experimentally synthesized [17-20].

Eu, one of the most reactive rare earth elements and an electronically strongly-correlated metal [21,22], adopts divalent state at ambient condition with a strong local magnetic moment [23]. At high pressure, it is anticipated to show trivalent state and a weak Van Vleck paramagnet, as more *f* electrons are squeezed into the *s*-, *p*-like electron, which might lead to a superconducting state [23-25]. In the H-rich conditions, Eu can accept hydrogen to form dihydrides or even trihydrides below 10 GPa [26,27]. Our previous simulations predicted that Eu reacts with hydrogen to form clathrate superhydrides EuH$_6$, EuH$_9$, and EuH$_{10}$ at high pressures, but superconductivity is yet to be predicted since theory has a difficult to give a good account on the electron-phonon coupling parameter of such a strong-correlated system [9]. This leaves an open question on the superconductivity of Eu superhydrides.

In this work, we target on the synthesis of Eu superhydrides under high pressure conditions using laser-heating diamond-anvil cell technique via a mixture of Eu and ammonia borane in an aim to find a clathrate superhydride that exhibits a strongly-correlated electronic property in nature, allowing us for subsequent investigation on how strongly-correlated effect in electronic structure can affect the superconductivity of superhydrides. Encouragingly, we did synthesize a series of Eu superhydrides of EuH$_3$, EuH$_5$, EuH$_6$ and EuH$_9$ in the pressure range of 80 - 170 GPa. Among all the superhydrides, the earlier anticipated clathrate EuH$_6$ and EuH$_9$ were successfully synthesized. EuH$_6$ is the second synthesized clathrate example following YH$_6$ thus far for a REH$_6$ stoichometry, but the first clathrate example for a strongly-correlated system in REH$_6$. The large H-derived electronic density of states at the Fermi level in clathrate EuH$_6$ implies the high potential of a good superconductor.

**RESULTS AND DISCUSSION**

In our previous study [9], the convex hull of Eu hydrides was constructed through density functional total-energy calculations by substituting Eu atom into known structures of REH$_3$, REH$_4$, REH$_6$, REH$_9$, and REH$_{10}$ (RE=La, Ce and Pr). It was found that EuH$_4$ and clathrate EuH$_9$ are stable compounds, while clathrate EuH$_6$ and EuH$_{10}$ lie above the convex hull in the whole pressure range of 100-400 GPa. Motivated by the theoretical results, we prepared four DACs named as samples C1, C2, C3, and C4, where a 2-μm thick sample of Eu was sandwiched between two BH$_3$NH$_3$ layers in a c-BN sample chamber. All samples are pressurized to 80-170 GPa in order to synthesize the clathrate superhydrides.

In sample C1, the pressure was loaded to 80 GPa at room temperature and then heated to 1,400 K by laser. The measured XRD pattern was plotted in Fig. 1a. We found that the resultant products were dominated by an fcc lattice, which could be identified as EuH$_2$ or EuH$_3$. However, it is difficult to distinguish them based on the current experimental data since their difference in lattice volume is too small. We then performed the enthalpy calculation and found that EuH$_3$ has a much lower enthalpy of 0.557 eV/atom than that of EuH$_2$+1/2H$_2$ when H$_2$ is in an excessive environment. From such an energy consideration, the fcc phase was thus identified as EuH$_3$. It is noteworthy that EuH$_3$ with a cubic close packing of Eu atom is isostructural to the ambient structure of LaH$_3$ [28], where each Eu atom is coordinated by 8 H atoms (Fig. 1b). The shortest H-H distance is 2.01 Å at 80 GPa.

After the first heating, the sample was further compressed to 130 GPa and then heated to ~1,600 K. The integrated XRD pattern is shown in Fig. 1c. A new phase was evident and considered to be a cubic lattice which is similar to the structure of *β*-UH$_3$ [29]. But from EOS comparison, we found the experimental volume of the new phase is much larger than the calculated volume of EuH$_3$ structure (Fig. S1), therefore, we considered the presence of additional hydrogen in same metal sublattice with that of *β*-UH$_3$. Structure searches of stoichiometric EuH$_4$, EuH$_5$ and EuH$_6$ at 150 GPa with fixing the position of Eu atoms were performed using the swarm-intelligence based CALYPSO structure prediction method [30,31]. We predicted a metastable *Pm-3n* structure of EuH$_5$ (Fig. 1d), the EOS of which is consistent with experimental P-V data (Fig. S1). This structure contains two inequivalent positions of Eu: one Eu atom in position I is surrounded by 12 H atoms at the corners of an icosahedron and another Eu atom in position II is surrounded by 6 H atoms forming a hexahedron. H$_2$ quasimolecules are evenly distributed in the interspace of icosahedrons.

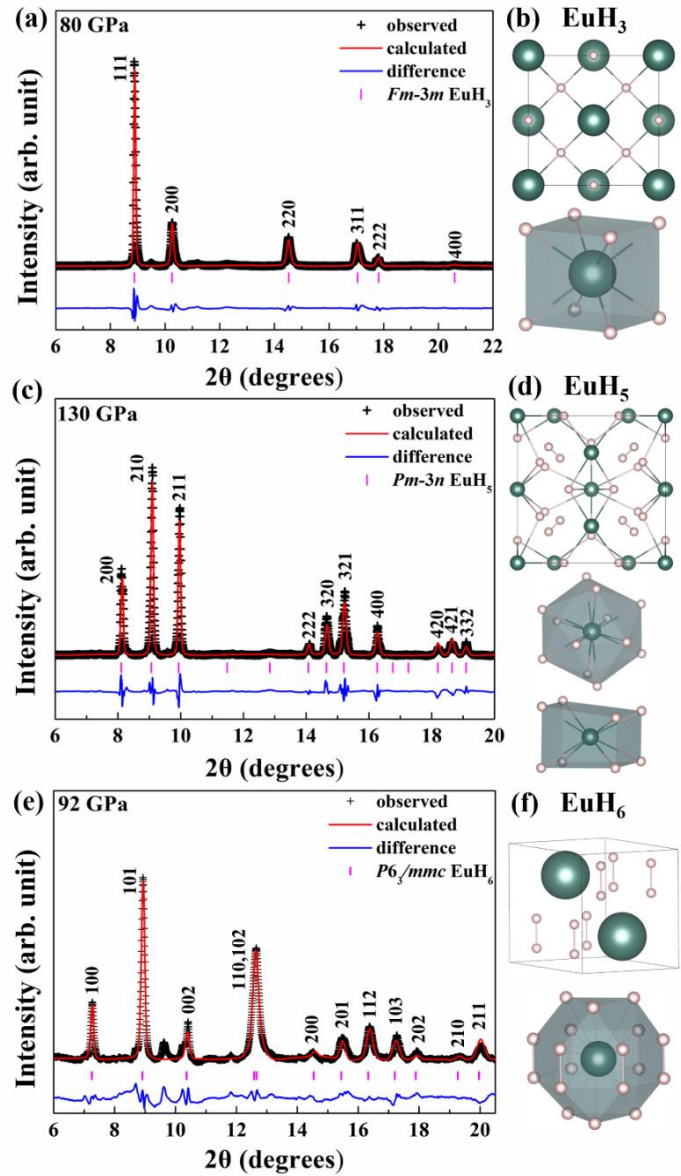

**Fig. 1.** Rietveld refinement of the experimental XRD pattern (left panel) and crystal structures (right panel) of (a),(b) EuH$_3$ at 80 GPa; (c),(d) EuH$_5$ at 130 GPa and (e),(f) non-clathrate EuH$_6$ at 92 GPa. Large and small balls represent the Eu atoms and H atoms, respectively.

In sample C2, the initial pressure was loaded up to 92 GPa and heated to a temperature of 2,100 K. The XRD patterns (Fig. 1e) obtained indicate an unexpected hexagonal $EuH_6$ which is a non-clathrate structure and was first theoretically reported in $ScH_6$[32], consisting of H-sharing 12-fold $EuH_{12}$ octahedrons (Fig. 1f). In this structure, hydrogen takes a quasimolecular state with d(H-H) = 1.143 Å.

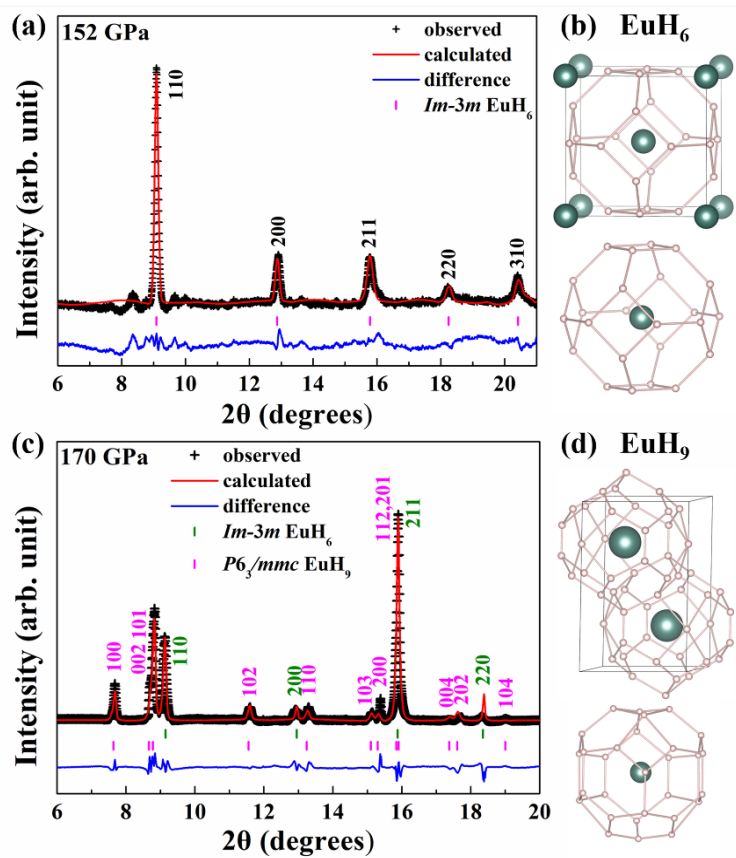

**Fig. 2.** Rietveld refinement of the experimental XRD pattern (left panel) and crystal structures (right panel) of (a),(b) clathrate $EuH_6$ at 152 GPa, and (c),(d) the mixture of clathrate $EuH_6$ and $EuH_9$ at 170 GPa. Large and small balls represent the Eu atoms and H atoms, respectively.

In order to synthesize the targeted clathrate structures, the sample C3 was directly compressed to an ultrahigh pressure of 152 GPa and laser heated to a temperature of 1,700 K. The measured XRD patterns of the sample (Fig. 2a) match the sodalite-like clathrate $EuH_6$ predicted in Ref. [9]. In the structure, each Eu atom is surrounded by 24 H atoms forming a $H_{24}$ cage and each cage is composed of 6 squares and 8 hexagons (Fig. 2b). The H-H distance is 1.3 Å and the nearest Eu–H distance is 2.051 Å at 152 GPa. This structure was first predicted in $CaH_6$ [1], but first synthesized in $YH_6$ [15,16]. $EuH_6$ is the second synthesized example and the first hexahydride synthesized in lanthanide hydrides.

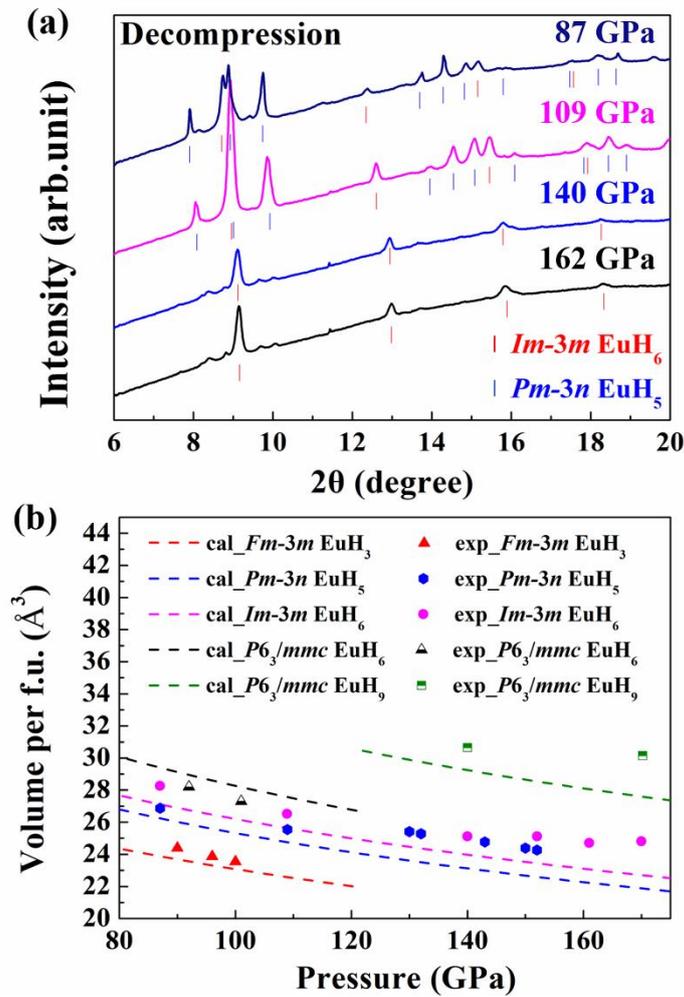

**Fig. 3. (a)** Experimental XRD patterns during decompression of sample C3 in the pressure range of 162–87 GPa. **(b)** The comparison of fitting EOS of all stoichiometries with the experimental P-V data. The dashed curves represent the calculated EOS. The symbols represent the experimental result.

During the release of pressure, the clathrate $EuH_6$ can be stabilized to at least 87 GPa, at a lower pressure which gradually decomposes into $EuH_5$ phase and $H_2$ (Fig. 3a). Electronic band structure and DOS calculations with a U parameter of 7.0 eV demonstrate that clathrate $EuH_6$ exhibits a metallic property (Fig. S2) with a relatively high H-derived DOS (0.23 states/eV/f.u.). The high superconductivity of the clathrate superhydrides arises from H atoms, where the H electrons contribute substantially to the electron DOS at the Fermi level [1,7,9,10]. Thus, clathrate $EuH_6$ is a promising candidate for high-temperature superconductors. Furthermore, the *f*-electrons of Eu make a main contribution to the electronic DOS at the Fermi level. U value plays an important role in governing electronic structure (Fig. S3).

The sample C4 was heated to 2,800 K with a loaded pressure at 170 GPa, aiming to synthesize superhydrides with a higher hydrogen content. Analysis of the measured XRD patterns showed that besides $EuH_6$, the clathrate stoichiometry $EuH_9$ was successfully synthesized (Fig. 2c) at this condition. The weight fractions of the $EuH_6$ and $EuH_9$ phases in the mixture were calculated by Rietveld refinement and estimated to be 46.43% and 53.56% respectively. $EuH_9$ was found to be isostructural to the earlier synthesized $YH_9$ [15] (Fig. 2d), in which each metal atom is surrounded by 29 H atoms forming a $H_{29}$ cage [9] and each cage is composed of 6 irregular squares, 6 pentagons, and 6 hexagons. We calculated its electronic band structure and DOS and compared them with that of $YH_9$ [3]. It can be seen that $EuH_9$ is a metallic phase, and noticeably, the projected DOS of hydrogen at the Fermi level is only 0.12 states/eV/f.u., which is much less than that of $YH_9$ (0.25 states/eV/f.u.).

Finally, we fitted the EOS of all stoichimetries and compared them with the experimental pressure-volume data (Fig. 3b). It is found that the experimental unit cell parameters and volumes are in good agreement with the theoretical data, which give further support on the valid structures and stoichiometries we identified from XRD data.

**CONCLUSION**

In summary, we have successfully synthesized a series of europium superhydrides, including $EuH_3$, $EuH_5$, $EuH_6$, and $EuH_9$ in the pressure range of 80 - 170 GPa. Among these hydrides, $EuH_6$ and $EuH_9$ have the unique sodalite-like clathrate structure with atomic-like hydrogen sublattice surrounding Eu atoms. Two non-clathrate structured phases of $EuH_5$ and $EuH_6$ are reported here for the prototype structure models in the system of lanthanide superhydrides. The calculated large H-derived DOS of clathrate $EuH_6$ implies its potential as a good candidate of high-$T_c$ superconductor. Theoretical simulations indicate that DFT+U correction has an essential impact on the calculated electronic property for the Eu-H system. This work paves a way for further experimental investigations on the superconductivity of superhydrides with a strongly-correlated effect in the electronic structure.

## METHODS

The experiments in the present work were conducted using laser-heated diamond anvil cell (DAC) techniques. The diamonds used in DACs had a culet with a diameter of 60-80 μm and were beveled at 8º to a diameter of about 280 μm. Europium hydrides were synthesized via a reaction of Eu (Alfa Aesar 99.99% purity) and $BH_3NH_3$ (Sigma-Aldrich 97% purity) at high pressure and high temperature conditions. The use of $BH_3NH_3$ as a source of $H_2$ has been demonstrated to be reliable by previous excellent results [13,16,33]. Composite gaskets consisting of a rhenium outer annulus and a mixture of cubic boron nitride (c-BN) with epoxy insert were employed to encapsulate the sample while isolating the thermal conductivity. Sample preparation and initial loading of the anvils were done in an inert Ar atmosphere (less than 0.01 ppm of oxygen and water) in a glove box to guarantee that the sample was properly isolated from the surrounding atmosphere. Afterwards, the DAC was taken out and the sample was compressed to the target pressure at room temperature. The pressure was determined from the shift in the high-frequency edge of the Raman spectrum gathered from the stressed tip of the diamond anvil [34].

In situ XRD data presented in this work were collected at the BL10XU beamline at the Spring-8 facility (Hyogo, Japan) with a wavelength of 0.4136 Å, and the X-ray spot size was around 3 μm × 2 μm. An imaging plate detector (RAXIS-IV; Rigaku) was used to collect the angle-dispersive XRD data. Primary processing and integration of the powder patterns were carried out using the Dioptas software [35]. Part of preliminary XRD measurements were also performed at the Shanghai Synchrotron Radiation Facility Beamline BL15U1 and Beijing Synchrotron Radiation Facility HP-Station 4W2. The Rietveld refinements were done using GSAS and EXPGUI packages [36]. The laser-heating experiments were performed by a two-sided SPI fiber laser with 1050 nm at BL10XU in Spring 8, and the temperature was determined by fitting the emission spectra from the surface of the heated sample to Planck's radiation law. All the crystal structure information obtained in this work is summarized in Table S1.

The equation of states (EOS) of $EuH_2$, $EuH_3$, $EuH_6$ and $EuH_9$ phases were calculated using density functional theory [37,38] within the generalized gradient approximation (Perdew–Burke–Ernzerhof functional) [39], and the projector-augmented wave method [40,41] as implemented in the VASP code [42-44]. The electron-ion interaction was described with the $5s^2 6s^2 5p^6 4f^7$ and $1s^1$ configurations treated as valence electrons for Eu and H, respectively. To ensure that all enthalpy calculations were well converged to about 1 meV/atom, the Brillouin zone was sampled using Γ-centered k-points meshes with a sufficient density ($2\pi \times 0.03$ Å$^{-1}$) in reciprocal space, as well as a kinetic energy cutoff of 800 eV. Eu has a half-filled *f*-shell, therefore the on-site Coulomb interactions are described by using DFT+U method with U = 7.0 eV [45]. The dependences of the volume on pressure were fitted by the 3rd order Birch–Murnaghan equation [46] to determine the main parameters of the EOS.

**Author contributions**

Liang Ma and Guangtao Liu equally contributed to this work.

**Competing interests**

The authors declare no competing interests.

**Acknowledgement**

This work is supported by the National Natural Science Foundation of China (Grant No. 11874175, 11974134), Science Challenge Project No. TZ2016001, the Program for JLU Science and Technology Innovative Research Team (JLUSTIRT), and National Key Research and Development Program of China (under Grant No. 2016YFB0201200 and 2016YFB0201201, 2016YFB0201204). We used the computing facilities at the High-Performance Computing Centre of Jilin University and Tianhe2-JK at the Beijing Computational Science Research Centre. XRD measurement were performed at BL10XU/Spring-8, Shanghai Synchrotron Radiation Facility Beamline BL15U1 and Beijing Synchrotron Radiation Facility HP-Station 4W2.

# Supporting Information

# Experimental Syntheses of Sodalite-like Clathrate $EuH_6$ and $EuH_9$ at Extreme Pressures

Liang Ma,[1,2] Guangtao Liu,[2] Yingying Wang,[2] Mi Zhou,[2] Hanyu Liu,[2] Feng Peng,[3] Hongbo Wang,[1,2*] and Yanming Ma[1,2,4†]

[1]State Key Laboratory of Superhard Materials, College of Physics, Jilin University, Changchun 130012, China

[2]International Center of Computational Method & Software, College of Physics, Jilin University, Changchun 130012, China

[3]College of Physics and Electronic Information, Luoyang Normal University, Luoyang 471022, China

[4]International Center of Future Science, Jilin University, Changchun 130012, China

**Corresponding authors:**
*whb2477@jlu.edu.cn, †mym@jlu.edu.cn

**Table S1.** Crystal structure of experimental synthesized Eu-H phases.

| Phase | Pressure, GPa | Lattice parameters | Coordinates | | | |
|---|---|---|---|---|---|---|
| $P6_3/mmc$ EuH$_9$ | 170 | a = b = 3.586 Å<br>c = 5.574 Å<br>α =β= 90°<br>γ = 120 | Eu (2c) | 0.33333 | 0.66667 | 0.25000 |
| | | | H (12k) | 0.15662 | 0.84338 | 0.55338 |
| | | | H (4f) | 0.33333 | 0.66667 | 0.84001 |
| | | | H (2b) | 0.00000 | 0.00000 | 0.75000 |
| $Im$-$3m$ EuH$_6$ | 152 | a = 3.689 Å<br>α =β=γ= 90° | Eu (2a) | 0.00000 | 0.00000 | 0.00000 |
| | | | H (12d) | 0.00000 | 0.50000 | 0.75000 |
| $P6_3/mmc$ EuH$_6$ | 92 | a = b = 3.774 Å<br>c = 4.577 Å<br>α =β= 90°<br>γ = 120 | Eu (2d) | 0.33333 | 0.66667 | 0.75000 |
| | | | H (12k) | 0.83991 | 0.16009 | 0.87527 |
| $Pm$-$3n$ EuH$_5$ | 130 | a = 5.846 Å<br>α =β=γ= 90° | Eu (6c) | 0.25000 | 0.00000 | 0.50000 |
| | | | Eu (2a) | 0.00000 | 0.00000 | 0.00000 |
| | | | H (24k) | 0.00000 | 0.15879 | 0.68672 |
| | | | H (16i) | 0.70243 | 0.70243 | 0.70243 |
| $Fm$-$3m$ EuH$_3$ | 80 | a = 4.626 Å<br>α =β=γ= 90° | Eu (6c) | 0.00000 | 0.00000 | 0.00000 |
| | | | H (2b) | 0.25000 | 0.25000 | 0.25000 |
| | | | H (2b) | 0.00000 | 0.00000 | 0.50000 |

**Fig. S1** Comparison of theoretical and experimental EOS of EuH$_3$ and EuH$_5$ in *Pm*-3*n* structure.

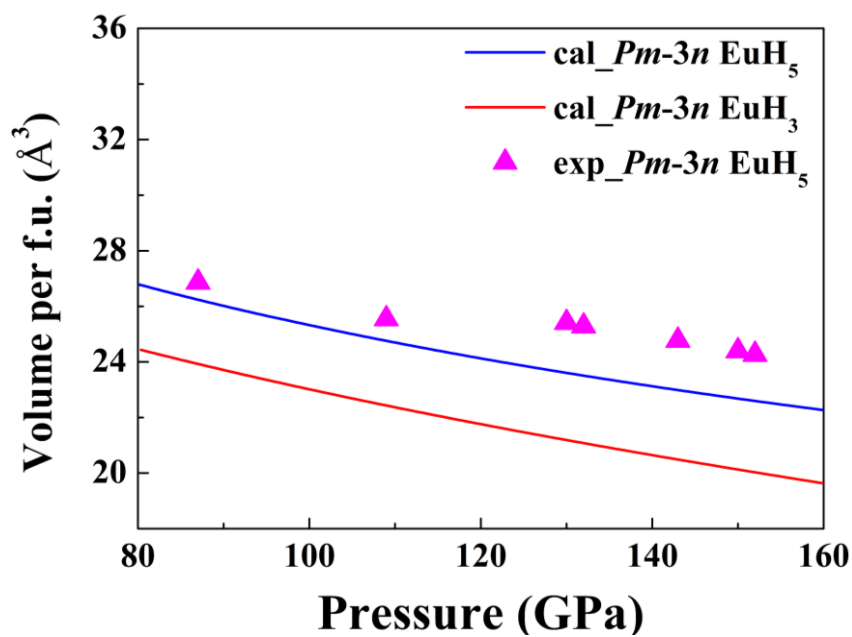

**Fig. S2.** Calculated band structure (left panel) and DOS (right panel) of (a) *Im*-3*m* EuH$_6$ at 150 GPa and (b) *P*6$_3$/*mmc* EuH$_9$ at 150 GPa. The insets show H-derived DOS.

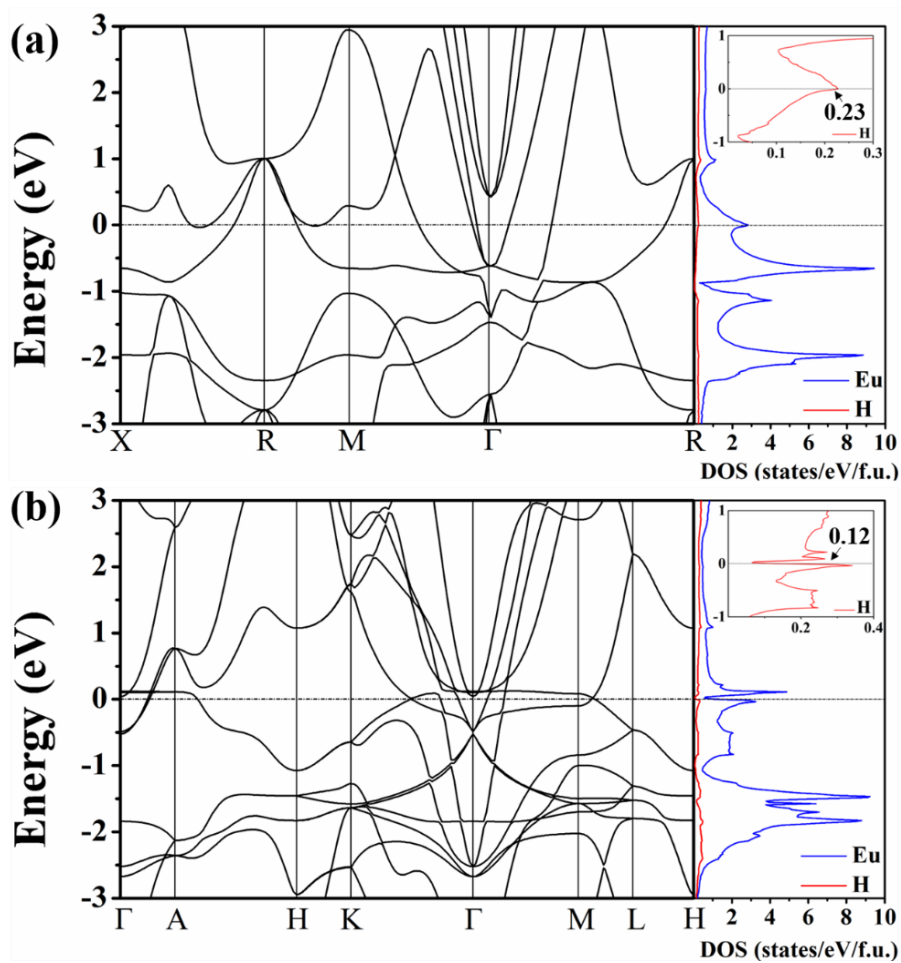

**Fig. S3** H-derived DOS of clathrate EuH₆ with and without U and the comparison with CaH₆ and YH₆.

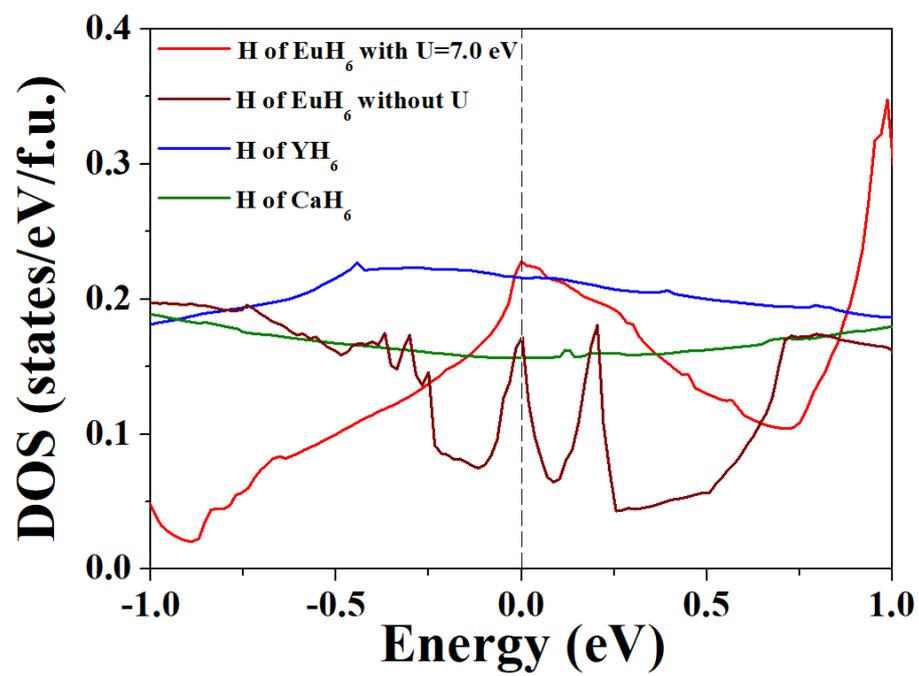